\documentclass[12pt,a4paper]{article}

\usepackage{epsfig}

\newcommand{\rii}{\mathrm{I\!I}}
\newcommand{\riii}{\mathrm{I\!I\!I}}
\newcommand{\riv}{\mathrm{I\!V}}
\newcommand{\rv}{\mathrm{V}}
\newcommand{\rvi}{\mathrm{V\!I}}

\begin{document}
\vspace*{-3cm}
\begin{flushright}
FISIST/28--2002/CFIF \\
hep-ph/0212057 \\
December 2002
\end{flushright}
\vspace{0.5cm}
\begin{center}
\begin{Large}
{\bf Hierarchy plus anarchy in quark \\[0.2cm] masses and mixings}
\end{Large}

\vspace{0.5cm}
J. A. Aguilar--Saavedra \\
{\it Departamento de F\'{\i}sica and Grupo de F\'{\i}sica de
Part\'{\i}culas (GFP), \\
Instituto Superior T\'ecnico, P-1049-001 Lisboa, Portugal}
\end{center}

\begin{abstract}
We introduce a new parameterisation of the effect of unknown corrections from
new physics on quark and lepton mass matrices. This parameterisation is used in
order to study how the hierarchies of quark masses and mixing angles are
modified by random perturbations of the Yukawa matrices. We discuss several
examples of flavour relations predicted by
different textures, analysing how these relations are influenced by the random
perturbations. We also comment on the unlikely possibility that unknown
corrections contribute significantly to the hierarchy of masses and mixings.
\end{abstract}

\section{Introduction}
\label{sec:1}

The observed quark masses and Cabibbo-Kobayashi-Maskawa (CKM)
mixing angles show a hierarchy
$m_t \gg m_c \gg m_u$, $m_b \gg m_s \gg m_d$, $|V_{us}| \gg |V_{cb}| \gg
|V_{ub}|$ that points to an underlying structure in the Yukawa couplings
of up and down-type quarks to the Higgs boson. This fact has motivated the
introduction of several models of mass matrices
trying to reproduce the experimental values of the quark masses and CKM
matrix elements using well-structured Yukawa patterns. In general, these models
consider that at very high energies, and for some reason unknown at present,
a weak basis is privileged among the infinite set of equivalent weak bases
related by rotations of the quark fields in flavour space. In this 
privileged basis, some flavour symmetry reduces the number of free parameters
in the Yukawa matrices, leading
to relations among masses and mixing angles that can be experimentally
tested. Examples of these models include patterns with texture zeros
\cite{papiro1}, democratic Yukawa couplings \cite{papiro2} or other symmetries
(see for instance \cite{papiro15,papiro15b}).

At lower scales, where the underlying flavour symmetry is broken, the
Yukawa textures are modified by renormalisation group (RG)
evolution and other radiative corrections, with several
contributions from Standard Model (SM) and new physics. These corrections are
unknown, and constitute what is called ``anarchy''. The ``anarchy'' may in
principle lead to modifications in the predictions of the original patterns.
The stability of these predictions, when the new contributions are included,
can be estimated with the addition of random perturbations to the Yukawa
matrices in the Lagrangian. Of course, the new physics contributions are not
random in nature, and could be computed if the complete theory was known.
At any rate, a simulation with random perturbations shows which quantities are
stable and which are not, and to what extent they are modified in this
case. It is already known that the strong hierarchy of masses and
mixings is not likely to be originated from ``anarchy'' \cite{papiro16}.
Recently, it has been argued \cite{papiro11} that in the quark sector the
possible contributions of the latter must be very small, in order to
preserve a hierarchical structure. However, this result has been obtained
for a particular texture of Yukawa couplings and in a particular weak basis. It
would be desirable to obtain a basis and model-independent result indicating,
at least qualitatively, how
the ``anarchy'' could change the hierarchy of quark masses and mixings. 

In the lepton sector the situation is rather different.
The masses of the charged leptons have been precisely measured, but
the knowledge about neutrino masses and mixings is still rather poor. After the
measurement of atmospheric neutrino oscillations, with
$\sin^2 \! \theta_\mathrm{atm} \geq 0.92$,
$\Delta m^2_\mathrm{atm} = (1.6-3.9) \times 10^{-3}$ GeV \cite{papiro17},
it is clear that the pattern of lepton masses and mixings is completely
different from the one in the
quark sector. In this direction, it has been suggested that the neutrino mass
matrices might be generated just from ``anarchy'' \cite{papiro3} (see also
Ref.~\cite{papiro3b}).
A complete analysis is not yet possible in the lepton sector,
for several reasons: ({\em i\/}) some of the mixing
angles have not still been measured; ({\em ii\/}) the neutrino masses remain
unknown and
only the differences of squared masses can be extracted from oscillation data;
({\em iii\/}) the Dirac or Majorana nature of the neutrinos is yet to be
determined, and in the first case Majorana-type mass terms are not present.

Our aim here is to study how the ``anarchy'' would affect a hierarchical
pattern, as the one found in the quark sector.
We first derive the formal structure of the unknown corrections to
the mass matrices, both for the case of quarks and leptons. We use this
formalism to investigate how random perturbations
modify a given hierarchy of quark masses and mixings. Then we discuss the
effect in some relations involving masses and CKM matrix
elements, which are predicted in several models of Yukawa patterns existing in
the literature. Finally, we explore the possibility of
a significative enhancement of the hierarchy of quark masses due to the
``anarchy''.

\section{Anarchy in the mass matrices}
\label{sec:2}

The starting point of our argument will be the observation that physical
quantities cannot depend on the weak basis chosen for the quark fields.
In the SM, the mass terms of the Lagrangian are written as
\begin{equation}
\mathcal{L}_m = -\bar u_L v\,Y^u u_R - \bar d_L v\,Y^d d_R + \mathrm{h.c.}\,,
\label{ec:1}
\end{equation}
with $v=174$ GeV the vacuum expectation value (VEV) of the Higgs boson, and
$Y^u$, $Y^d$ the Yukawa matrices, of dimension $3 \times 3$ in flavour space. 
In some SM extensions, for instance in the minimal supersymmetric Standard
Model (MSSM), the masses are originated from the VEV's of two Higgs bosons,
$v_u$ and $v_d$, with $v_u^2+v_d^2 = v^2$. Defining $\beta$ by
$\tan \beta = v_u/v_d$, the mass matrices for up and down quarks are in this
case $v \sin \beta \, Y^u$ and $v \cos \beta \, Y^d$, respectively, with an
extra $\beta$-dependent factor. We later comment on how this possibility
modifies our analysis.

It is well known that under a change of basis
\begin{equation}
\left( \! \begin{array}{c}u_L \\ d_L \end{array} \! \right) =
U_L \left( \! \begin{array}{c}u'_L \\ d'_L \end{array} \! \right) ~~,~~~
u_R = U_R^u ~ u'_R ~~,~~~ d_R = U_R^d ~ d'_R 
\label{ec:2}
\end{equation}
the Yukawa matrices transform as
\begin{equation}
Y^u \to U_L^\dagger \, Y^u \, U_R^u ~~,~~~ Y^d \to U_L^\dagger \, Y^d \, U_R^d
\,.
\label{ec:3}
\end{equation}
Let us assume that some perturbations are added to the original Yukawa matrices,
\begin{equation}
Y^u \to Y^u + \delta Y^u ~~,~~~ Y^d \to Y^d + \delta Y^d \,.
\label{ec:4}
\end{equation}
The matrices $\delta Y^u$, $\delta Y^d$ are functions of $Y^u$, $Y^d$ and other
SM and new physics parameters.
Under the change of basis in Eqs.~(\ref{ec:3}), the perturbations
must transform as
\begin{equation}
\delta Y^u \to U_L^\dagger \, \delta Y^u \, U_R^u ~~,~~~
\delta Y^d \to U_L^\dagger \, \delta Y^d \, U_R^d
\,,
\label{ec:5}
\end{equation}
since the physical observables must be independent of the choice of weak basis
\footnote{These transformation properties for $\delta Y^u$ and $\delta Y^d$
do not assume that the Lagrangian is invariant under the transformations in
Eqs.~(\ref{ec:3}) alone. Within the SM, the Lagrangian is invariant under these
transformations, but this does not happen in some of its extensions, for
instance in the MSSM. Besides, at very high energies, some symmetry might
single out a special weak basis. Below that scale, and in particular at low
energies, this symmetry is broken.}.
These transformation laws imply that the perturbations have the form
\begin{eqnarray}
\delta Y^u & = & \lambda_u \, Y^u + \zeta_u \, Y^u Y^{u \dagger} Y^u
+ \eta_u \, Y^d Y^{d \dagger} Y^u + \cdots \,, \nonumber \\
\delta Y^d & = & \lambda_d \, Y^d + \zeta_d \, Y^d Y^{d \dagger} Y^d
+ \eta_d \, Y^u Y^{u \dagger} Y^d + \cdots \,.
\label{ec:6}
\end{eqnarray}
The dimensionless coefficients $\lambda_i$, $\zeta_i$, $\eta_i$ may be
functions of $Y^u$ and $Y^d$ that are invariant under
$\mathrm{SU}(3)$ flavour rotations, {\em e.g.}
$\mathrm{tr}\; Y^u Y^{u\dagger}$,
$\mathrm{tr}\; Y^d Y^{d\dagger}$, etc. and of other SM and new physics
parameters as well.
The higher-order terms are expected to be smaller (if perturbation theory is
valid), thus in the expansions of
Eqs.~(\ref{ec:6}) the products with five or more Yukawa matrices have been
omitted. The effect of the $\lambda_i$ terms in Eqs.~(\ref{ec:6}) is to rescale
the masses by common factors $(1+\lambda_u)$ for up quarks and
$(1+\lambda_d)$ for down quarks, without affecting the hierarchy and the
mixing. The $\zeta_i$ terms also rescale the masses, but with different factors
for each quark, $m_q \to m_q \, (1+\zeta_i \,m_q^2/v^2)$, and then they modify
the hierarchies. The $\eta_i$ terms are the
lowest-order ones that modify the CKM matrix.

In principle, in SM extensions there may exist a matrix $X$ (not necessarily
square) of couplings between the quarks and other particles, either transforming
on the left or on the right side as one of the Yukawa matrices.
For instance, if $X$
transforms under the change of basis in Eqs.~(\ref{ec:3}) as
\begin{equation}
X \to U_L^\dagger \, X \, U_R^X \,,
\label{ec:8}
\end{equation}
this matrix would originate terms $X X^\dagger Y^u$ and $X X^\dagger Y^d$ in
Eqs.~(\ref{ec:6}). However, if diagrams involving $X$ couplings gave significant
corrections to the Higgs Yukawa vertices with a new flavour structure,
analogous diagrams with a photon, gluon or $Z$ boson would give
similar contributions
to flavour-changing processes, which are experimentally very suppressed. Thus we
disregard this possibility in the following.

It is worthwhile noting that the formal structure of the perturbations to the
Yukawa matrices, derived here from weak-basis independence arguments, coincides
with the expression of the RG equations in the SM. Setting for instance
in Eqs.~(\ref{ec:6})
\begin{eqnarray}
\lambda_u & = & \frac{1}{16 \pi^2} \left \{
\mathrm{tr} \left[ 3 Y^u Y^{u\dagger} + 3 Y^d Y^{d \dagger} 
+ Y^e Y^{e \dagger} \right] -
\left( \frac{17}{20} g_1^2 + \frac{9}{4} g_2^2 + 8 g_3^2 \right)
\right \} \delta t \,, \nonumber  \\
\lambda_d & = & \frac{1}{16 \pi^2} \left \{
\mathrm{tr} \left[ 3 Y^u Y^{u\dagger} + 3 Y^d Y^{d \dagger} 
+ Y^e Y^{e \dagger} \right] -
\left( \frac{1}{4} g_1^2 + \frac{9}{4} g_2^2 + 8 g_3^2 \right)
\right \} \delta t \,, \nonumber  \\
\zeta_u & = & \frac{1}{16 \pi^2} \left\{ \frac{3}{2} \right\} \delta t 
\,, \nonumber \\
\zeta_d & = & \frac{1}{16 \pi^2} \left\{ \frac{3}{2} \right\} \delta t
\,, \nonumber \\
\eta_u & = & \frac{1}{16 \pi^2} \left\{ -\frac{3}{2} \right\} \delta t 
\,, \nonumber \\
\eta_d & = & \frac{1}{16 \pi^2} \left\{ -\frac{3}{2} \right\} \delta t
\,,
\label{ec:7}
\end{eqnarray}
the SM one-loop RG equations for the Yukawa couplings are recovered
\cite{papiro4,papiro4b}
\footnote{Note that our Yukawa matrices $Y$ correspond to
$Y^\dagger$ in Refs.~\cite{papiro4,papiro4b}.}.
In Eqs.~(\ref{ec:7}), $Y^e$ is the Yukawa
matrix for the charged leptons, $g_3$, $g_2$ and $g_1$ are the coupling
constants of the gauge group
$\mathrm{SU}(3) \otimes \mathrm{SU}(2) \otimes \mathrm{U}(1)$ and
$t = \log \mu$ the logarithm of the renormalisation scale.

In the lepton sector the analysis is more involved, because of the possible
presence of Majorana fermions. In this case the mass terms of the Lagrangian
read
\begin{equation}
\mathcal{L}_m = - \bar e_L v \, Y^e e_R 
- \frac{1}{2} \left( \bar \nu_L ~ \overline {\nu_R^c} \right)
\left( \begin{array}{cc} M_L & v \, Y^\nu \\ v \, Y^{\nu T} & M_R \end{array}
\right)
\left( \begin{array}{c} \nu_L^c \\ \nu_R \end{array} \right) \,.
\label{ec:9}
\end{equation}
The Dirac mass matrices $v \, Y^e$ and $v \, Y^\nu$ arise from Yukawa couplings
to the Higgs boson. The Majorana mass matrix $M_R$ can be included as a bare
term in the Lagrangian, because the right-handed neutrinos are singlets under
the gauge group. The Majorana mass term $M_L$ involving the left-handed neutrino
fields can only exist at tree-level in a renormalisable Lagrangian if a Higgs
triplet is present (this is practically excluded by precision electroweak data),
and $M_L$ is usually set to zero. Both $M_L$ and $M_R$ are symmetric
matrices. We define $M_L \equiv \Lambda_L \hat M_L$, $M_R \equiv \Lambda_R \hat
M_R$, with $\Lambda_L$ and $\Lambda_R$ constants with the dimension of mass, in
order to express $M_L$ and $M_R$ in terms of dimensionless matrices.

Under a change of weak basis
\begin{equation}
\left( \! \begin{array}{c}\nu_L \\ e_L \end{array} \! \right) =
V_L \left( \! \begin{array}{c}\nu'_L \\ e'_L \end{array} \! \right) ~~,~~~
\nu_R = V_R^\nu ~ \nu'_R ~~,~~~ e_R = V_R^e ~ e'_R 
\label{ec:10}
\end{equation}
we have
\begin{eqnarray}
Y^e & \to & V_L^\dagger \, Y^e \, V_R^e ~~,~~~
Y^\nu \to V_L^\dagger \, Y^\nu \, V_R^\nu \,, \nonumber \\
\hat M_L & \to & V_L^\dagger \hat M_L V_L^* ~~,~~~
\hat M_R \to V_R^{\nu T} \hat M_R V_R^\nu \,.
\label{ec:11}
\end{eqnarray}
The computation of all the products up to order three transforming as $Y^e$,
$Y^\nu$, $\hat M_L$ and $\hat M_R$ can be easily done with the programs in
Ref.~\cite{papiro5}. It is found that the perturbations to these matrices can
be expanded as
\begin{eqnarray}
\delta Y^e & = & \lambda_e \, Y^e
+ \zeta_e \, Y^e Y^{e \dagger} Y^e
+ \eta_{e1} \, Y^\nu Y^{\nu \dagger} Y^e
+ \eta_{e2} \, \hat M_L \hat M_L^\dagger Y^e
+ \cdots \,, \nonumber \\
\delta Y^\nu & = & \lambda_\nu \, Y^\nu + \zeta_\nu Y^\nu Y^{\nu \dagger} Y^\nu
+ \eta_{\nu 1} \, Y^e Y^{e \dagger} Y^\nu
+ \eta_{\nu 2} \, \hat M_L \hat M_L^\dagger Y^\nu \nonumber \\
& & + \xi_{\nu 1} \, Y^\nu \hat M_R^\dagger \hat M_R
+ \xi_{\nu 2} \, \hat M_L Y^{\nu *} \hat M_R
+ \cdots \,, \nonumber \\
\delta \hat M_L & = & \lambda_L \, \hat M_L 
+ \zeta_L \, \hat M_L \hat M_L^\dagger \hat M_L
+ \eta_{L1} \left( Y^e Y^{e \dagger} \hat M_L 
  + \hat M_L Y^{e*} Y^{eT} \right) \nonumber \\
& & + \eta_{L2} \left( Y^\nu Y^{\nu \dagger} \hat M_L
  + \hat M_L Y^{\nu *} Y^{\nu T} \right)
+ \chi_L \, Y^\nu \hat M_R^\dagger Y^{\nu T}
+ \cdots \,, \nonumber \\
\delta \hat M_R & = & \lambda_R \, \hat M_R
+ \zeta_R \, \hat M_R \hat M_R^\dagger \hat M_R
+ \eta_R \left( M_R Y^{\nu \dagger} Y^\nu
  + Y^{\nu T} Y^{\nu *} M_R \right) \nonumber \\
& & + \chi_R \, Y^{\nu T} \hat M_L^\dagger Y^\nu
+ \cdots \,,
\label{ec:12}
\end{eqnarray}
where the $\lambda$, $\zeta$, $\eta$, $\xi$, $\chi$ coefficients are functions
of $Y^e$, $Y^\nu$, $\hat M_L$, $\hat M_R$ and of the coupling constants,
invariant under the transformations of Eqs.~(\ref{ec:11}).
Under additional assumptions, the number of terms can be
reduced. For instance, assuming that $\hat M_L = 0$ at tree-level,
the terms with $\eta_{e2}$, $\eta_{\nu 2}$, $\xi_{\nu 2}$, $\lambda_L$,
$\zeta_L$, $\eta_{L1}$, $\eta_{L2}$ and $\chi_R$
can be dropped from the expressions.
Notice that even setting $\hat M_L = 0$ in the Lagrangian, our argument
allows a non-vanishing
$\delta \hat M_L = \chi_L Y^\nu \hat M_R^\dagger Y^{\nu T}$ to be generated,
though some symmetry in the Lagrangian may imply $\chi_L = 0$.
The analysis of the effects of the ``anarchy'' in the lepton sector
will be presented elsewhere \cite{papiro6}.

\section{Effects of anarchy in the hierarchy of masses and mixing angles}
\label{sec:3}

We first study how the ``anarchy'' may change a hierarchical structure in the
Yukawa couplings. For our discussion we take as benchmark the 
SM Yukawa matrices in the $\overline \mathrm{MS}$ scheme at the scale $M_Z$
(since the expressions in Eqs.~(\ref{ec:6}) are basis-independent, we can use
any basis for the evaluations). The values used for the quark masses are
collected
in Table~\ref{tab:1}. For the CKM matrix we use, in the standard
parameterisation \cite{papiro7}, $|V_{us}| = 0.2224$,
$|V_{ub}| = 0.00362$, $|V_{cb}| = 0.0402$ and $\delta = 1.014$.

\begin{table}[htb]
\begin{center}
\begin{tabular}{cccc}
\hline
\hline
$m_u$ & $0.0016$ \\
$m_c$ & $0.68$  \\
$m_t$ & $175.6$ \\
$m_d$ & $0.0033$ \\
$m_s$ & $0.067$ \\
$m_b$ & $2.9$ \\
\hline
\hline
\end{tabular}
\caption{Quark masses (in GeV) used in the numerical evaluations.
\label{tab:1}}
\end{center}
\end{table}

We use the perturbations to the Yukawa matrices given in Eqs.~(\ref{ec:6}),
with random coefficients $\lambda_i$, $\zeta_i$, $\eta_i$. The
moduli of these six independent parameters are generated with a Gaussian
distribution centred at zero, and for simplicity we assume that the standard
deviations coincide:
\begin{equation}
\langle |\lambda_i|^2 \rangle^\frac{1}{2} =
\langle |\zeta_i|^2 \rangle^\frac{1}{2} = 
\langle |\eta_i|^2 \rangle^\frac{1}{2} \equiv \kappa \,.
\label{ec:13}
\end{equation}
This is not a serious bias in the analysis, because the moduli of the random
parameters
are not fixed to be all equal, and only the standard deviations of the
distributions are assumed to be the same. The phases are generated uniformly
between $0$ and $2\pi$. The only effect of the $\lambda_i$ terms is to change
the ratio $m_b/m_t$; the mixing angles and the ratios
of masses of quarks of the same charge are not modified by them.
Here it is worth remarking that the situation when two different scalars give
masses to the up and down-type
quarks can be reproduced with the substitutions
\begin{equation}
Y^u \to Y^u \sin \beta ~,~~~ Y^d \to Y^d \cos \beta
\label{ec:13b}
\end{equation}
in Eqs.~(\ref{ec:6}). This is equivalent to considering
\begin{eqnarray}
\langle |\lambda_i|^2 \rangle^\frac{1}{2} & = & \kappa \,, \nonumber \\
\langle |\zeta_u|^2 \rangle^\frac{1}{2} =
\langle |\eta_d|^2 \rangle^\frac{1}{2} & = & \kappa \sin^2 \! \beta \,,
\nonumber \\
\langle |\zeta_d|^2 \rangle^\frac{1}{2} =
\langle |\eta_u|^2 \rangle^\frac{1}{2} & = & \kappa \cos^2 \! \beta
\label{ec:13c}
\end{eqnarray}
instead of Eqs.~(\ref{ec:13}). The effect of $\tan \beta$ is to modify
the standard deviations of the distributions of the random parameters.
The qualitative behaviour with two scalars is the same as in the case discussed
below, with only one Higgs. The quantitative behaviour is similar, as
long as $\tan \beta \sim 1$.

\begin{figure}[htb]
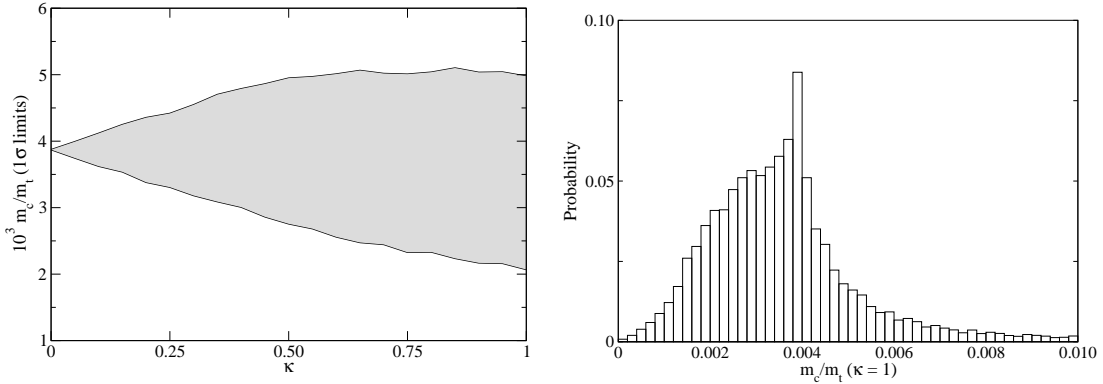

\begin{center}
\begin{tabular}{cc}
\mbox{\epsfig{file=Figs/mcmt.eps,width=7cm,clip=}} &
\raisebox{-0.13cm}{\epsfig{file=Figs/mcmt-k1.eps,width=7.11cm,clip=}}
\end{tabular}
\end{center}
\caption{Effect of the random perturbations on the ratio $m_c/m_t$.
\label{fig:mcmt} }
\end{figure}

In the study of the consequences of ``anarchy'' the following procedure is
applied:
we fix a value of $\kappa$ and generate a set of random matrices, with a number
of elements between 2000
(for $\kappa \simeq 0$) and 8000 (for $\kappa \simeq 1$). We then
select some quantity,
for instance the ratio $m_c/m_t$, and examine its distribution over the set.
The $1 \sigma$ limits on this quantity are defined as the
boundaries of the 68.7\% confidence level central interval, evaluated from the
sample of random matrices. We begin the discussion with this example of
$m_c/m_t$. We find that this quantity remains
fairly stable, even for
relatively large values of $\kappa$, {\em e.g.}
$\kappa = 1$. In Fig.~\ref{fig:mcmt} (left) we plot the
$1\sigma$ limits on $m_c/m_t$, for $\kappa$ between 0 and 1. In
Fig.~\ref{fig:mcmt} (right) we plot the distribution of the values of $m_c/m_t$
for $\kappa = 1$.

Several comments are in order:
\begin{enumerate}
\item The distribution of $m_c/m_t$ is peaked around the unperturbed value
$(m_c/m_t)_0 = 3.9 \times 10^{-3}$, because the random parameters are generated
with a Gaussian distribution centred at zero.
It spreads over values larger and smaller than $(m_c/m_t)_0$, but with larger
$\kappa$ the smaller values are favoured. This can be observed in both
plots, and evidences an average increase of $m_t$ with respect to $m_c$,
due to the larger Yukawa coupling of the former. The ratio $m_c/m_t$ is modified
mainly by the $\zeta_u$ term, giving corrections
$\delta m_c = \zeta_u \, m_c^3/v^2$, $\delta m_t = \zeta_u \, m_t^3/v^2$ that
are much larger for the top quark.

\item The spread of the values of $m_c/m_t$ increases with $\kappa$, as it can
be expected. The growth is linear for small $\kappa$ but it is attenuated for
$\kappa \geq 0.5$, as can be observed in Fig.~\ref{fig:mcmt} (left).
The upper limit on $m_c/m_t$ reaches a ``saturation'' value for $\kappa \geq
0.5$ and does not increase beyond this value.
This can be understood as follows: As the corrections
are larger for the top quark, the ratio $m_c/m_t$ will generically be smaller,
and the only chance to have a larger $m_c/m_t$ is with a cancellation
between $m_t$ and $\delta m_t$. For $\kappa$ sufficiently large, the probability
of this fine tuning to happen is practically constant.

\item  Most of the values of $m_c/m_t$ remain close to their original value.
Even for $\kappa = 1$, 68.7\% of the distribution is between the values
$2.0 \times 10^{-3}$ and $5.0 \times 10^{-3}$, very similar to $(m_c/m_t)_0$.
It is noticeable
the long tail of the distribution in Fig.~\ref{fig:mcmt} (right). For
$\kappa = 1$, the
largest value of $m_c/m_t$ found in the set is $0.15$, for which the mass
hierarchy is washed away. Such values are reached only in an extremely small
fraction of the sample.
\end{enumerate}
These plots must not be interpreted as providing any limit on the size of the
``anarchy'', on the basis of the experimental value of $m_c/m_t$. Instead, their
meaning is that the ratio $m_c/m_t$ is very stable under perturbations and
the mass hierarchy is maintained (due to points 2 and 3 above): from an original
value $(m_c/m_t)_0 = 3.9 \times 10^{-3}$ and with $\kappa = 1$ we obtain ratios
between $2.0 \times 10^{-3}$ and $5.0 \times 10^{-3}$ most of the time.

\begin{figure}[htb]
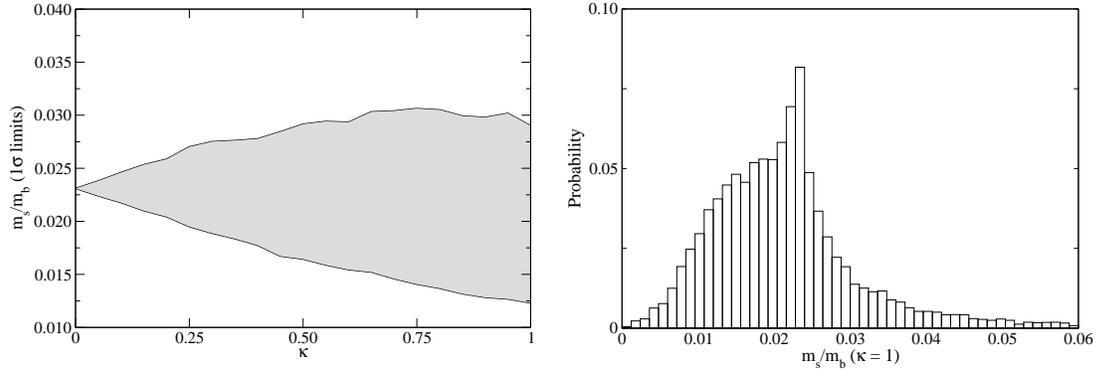

\begin{center}
\begin{tabular}{cc}
\mbox{\epsfig{file=Figs/msmb.eps,width=7cm,clip=}} &
\raisebox{-0.14cm}{\epsfig{file=Figs/msmb-k1.eps,width=7cm,clip=}}
\end{tabular}
\end{center}
\caption{Effect of the random perturbations on the ratio $m_s/m_b$.
\label{fig:msmb} }
\end{figure}

The effect of the random perturbations in the ratio $m_s/m_b$ is practically the
same, as can be observed in Fig.~\ref{fig:msmb}. Although with different
numerical values, this ratio shows the same behaviour under random perturbations
as $m_c/m_t$, and the above comments apply in this case as well.
The leading correction to $m_s/m_b$ is due to the $\eta_d$ term,
$\delta m_s \simeq \eta_d \, m_s m_c^2/v^2$,
$\delta m_b \simeq \eta_d \, m_b m_t^2/v^2$, and the effects are analogous.

\begin{figure}[htb]
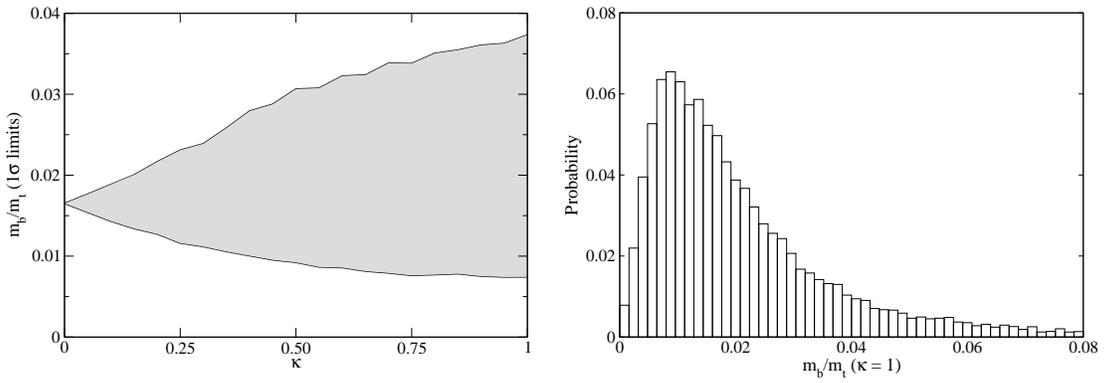

\begin{center}
\begin{tabular}{cc}
\mbox{\epsfig{file=Figs/mbmt.eps,width=7cm,clip=}} &
\raisebox{-0.13cm}{\epsfig{file=Figs/mbmt-k1.eps,width=7.11cm,clip=}}
\end{tabular}
\end{center}
\caption{Effect of the random perturbations on the ratio $m_b/m_t$.
\label{fig:mbmt} }
\end{figure}

The analysis of the two mass ratios $m_u/m_c$ and $m_d/m_s$ shows
that they do not change when random perturbations
are added to the Yukawa matrices, remaining at the values
$(m_u/m_c)_0 = 2.3 \times 10^{-3}$, $(m_d/m_s)_0 = 0.049$.
This is the same behaviour as under RG evolution \cite{papiro8},
where it is found that these ratios depend weakly on the scale. Here
we find examples
where $m_d/m_s$ raises to 0.10, but this only happens in a negligibly small
fraction of the sample. Therefore, the mass hierarchy between the
first and second
quark generations is also maintained. The remaining independent ratio $m_b/m_t$
exhibits a different behaviour, as can be readily noticed in
Fig.~\ref{fig:mbmt}. The maximum of the distribution is displaced to smaller
values
of $m_b/m_t$, and additionally there is a long tail for larger values of this
quantity. This can be understood as follows: Setting $|V_{tb}| = 1$,
Eqs.~(\ref{ec:6}) imply for the third generation
\begin{eqnarray}
\delta m_t & = & \lambda_u \, m_t + \zeta_u \, m_t^3/v^2
+ \eta_u \, m_t m_b^2/v^2 \,,
\nonumber \\
\delta m_b & = & \lambda_d \, m_b + \zeta_d \, m_b^3/v^2
+ \eta_d \, m_b m_t^2/v^2 \,.
\label{ec:14}
\end{eqnarray}
The enhancement for small values of $m_b/m_t$ is due to the large term
$\zeta_u \, m_t^3/v^2$, while the long tail is a consequence of the term
$\eta_d \, m_b m_t^2/v^2$, also important. The change of this ratio with
respect to
its original value is limited, though more pronounced than in the previous
cases.

\begin{figure}[htb]
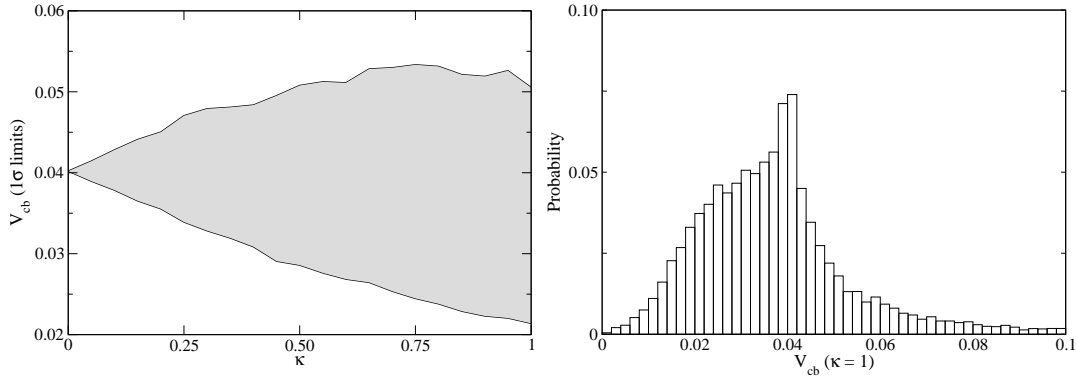

\begin{center}
\begin{tabular}{cc}
\mbox{\epsfig{file=Figs/Vcb.eps,width=7cm,clip=}}
\raisebox{-0.13cm}{\epsfig{file=Figs/Vcb-k1.eps,width=7.05cm,clip=}}
\end{tabular}
\end{center}
\caption{Effect of the random perturbations on $|V_{cb}|$.
\label{fig:Vcb} }
\end{figure}

The modulus of the CKM matrix element $V_{us}$ is not affected by the random
perturbations. As happens for the ratios $m_u/m_c$ and $m_d/m_s$, there are
fine-tuned values of
the random parameters for which the corrections yield a much larger
value, for instance $|V_{us}| \sim 0.4$. We stress that this only happens in an
insignificant fraction of the sample, and for the rest we find $|V_{us}|$ almost
fixed at its original value. The CKM phase $\delta$ does not change either.
The moduli of $V_{cb}$ and $V_{ub}$ change
when the perturbations are added, but their ratio remains constant. (Besides,
this is the same behaviour with RG evolution that is found for $V_{us}$,
$V_{cb}$, $V_{ub}$ and the phase $\delta$ \cite{papiro8}.)
The distribution of values of $|V_{cb}|$
can be seen in Fig.~\ref{fig:Vcb}, and coincides with the distribution of
$|V_{ub}|$ up to an overall factor.

Comparing these plots with Figs.~\ref{fig:mcmt} and \ref{fig:msmb} we observe
that the effect of the ``anarchy'' on the moduli of the mixing angles $V_{cb}$
and $V_{ub}$ and in the ratios of masses $m_c/m_t$, $m_s/m_b$, $m_u/m_t$ and
$m_d/m_b$ is similar.
This fact is due to the formal structure of the random perturbations,
and has non-trivial implications in some flavour relations in the
next Section.

\section{Effects in flavour relations}
\label{sec:4}

We have selected nine simple flavour relations predicted by several models in
the literature. Here we do not pretend to test whether these relations are
actually fulfilled by experimental data. Such analysis ought to be carried out
at a very high energy scale, of the order of the unification scale
$M_X \sim 10^{16}$ GeV (where a hypothetical flavour symmetry is
unbroken), using the Yukawa matrices at that scale. Instead, our aim is
to investigate whether these relations are modified by the ``anarchy''.
This analysis can be safely done at the scale $M_Z$.
We prefer to perform the simulations at this scale because
the extrapolation to the unification scale $M_X$ depends on the particle
content of the theory between $M_Z$ and $M_X$, and therefore it
is different in the SM and its extensions. For instance, in the MSSM the running
of some parameters, like $|V_{ub}|$, depends strongly on the parameter
$\tan \beta$.

\begin{figure}[htb]
\begin{center}
\begin{tabular}{cc}
\mbox{\epsfig{file=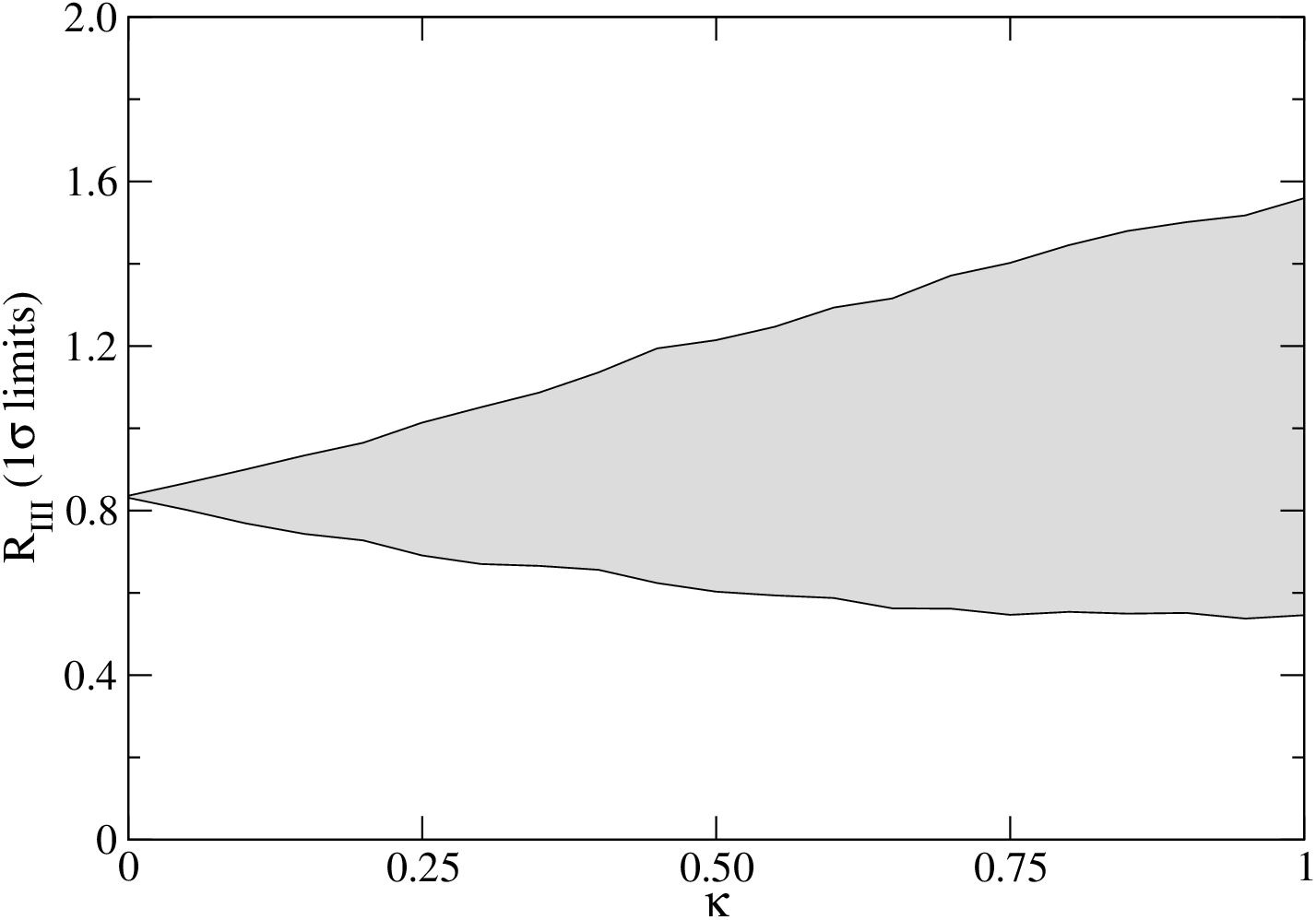,width=7cm,clip=}} &
\raisebox{-0.16cm}{\epsfig{file=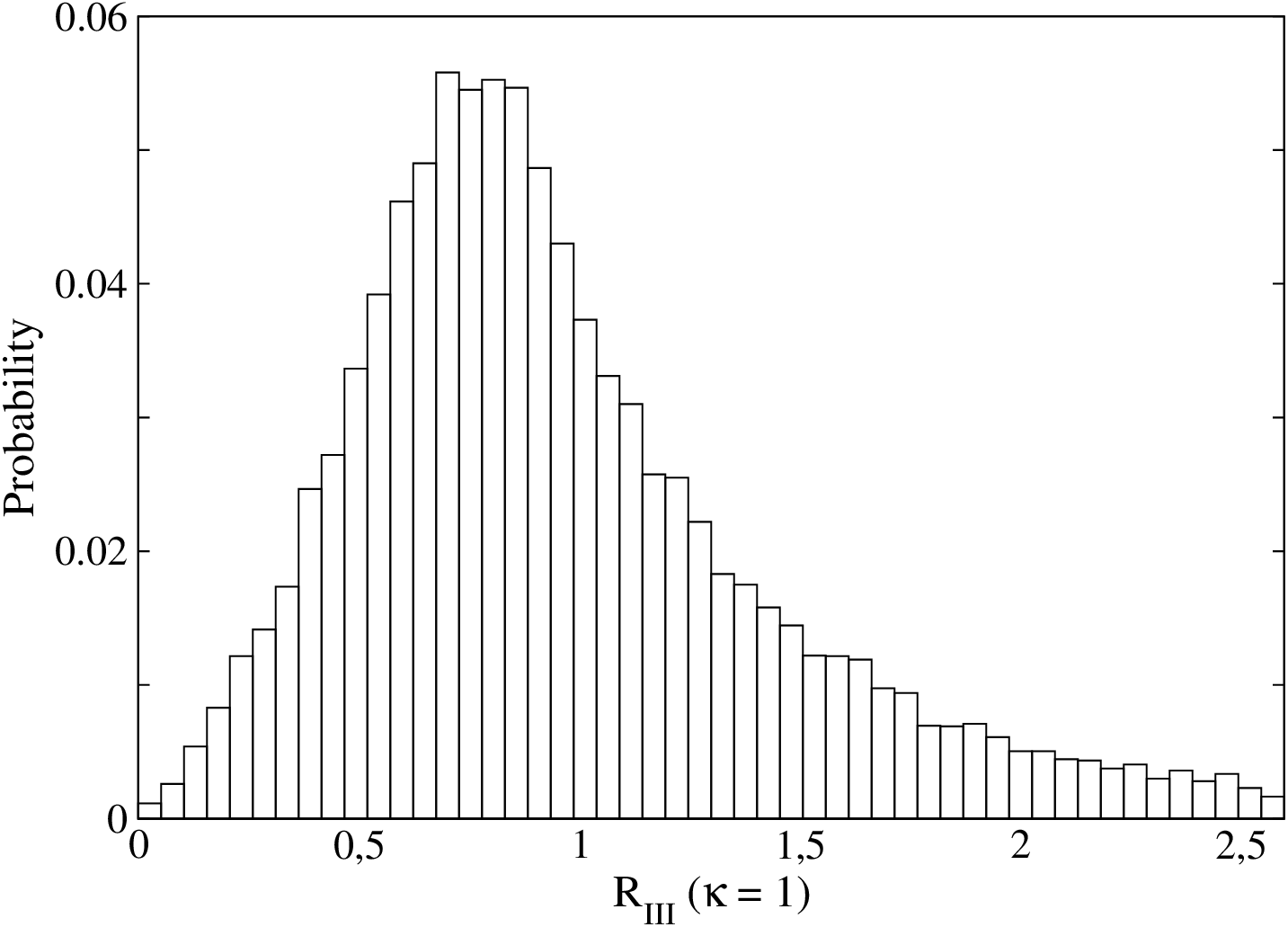,width=7.02cm,clip=}}
\end{tabular}
\end{center}
\caption{Effect of the random perturbations on the ratio $R_\riii$, defined in
the text.
\label{fig:R3} }
\end{figure}

Relation I is \cite{papiro9,papiro9b}
\begin{equation}
|V_{us}| = \sqrt \frac{m_d}{m_s} \,,
\label{ec:15}
\end{equation}
predicted for instance by the five textures proposed in Ref.~\cite{papiro1}.
There are additional small corrections to this equality that we ignore.
As neither
$|V_{us}|$ nor the ratio $m_d/m_s$ change under random perturbations,
the accuracy (or inaccuracy) of this relation is not modified. Relation II
is
\begin{equation}
\left| \frac{V_{ub}}{V_{cb}} \right| = \sqrt \frac{m_u}{m_c} \,.
\label{ec:16}
\end{equation}
It is predicted in textures 1, 2 and 4 of Ref.~\cite{papiro1}.
In this case, the ratios $V_{ub}/V_{cb}$ and $m_u/m_c$ do not change, and this
relation also remains unaffected. Relation III is
\begin{equation}
|V_{ub}| = \sqrt \frac{m_u}{m_t} \,,
\label{ec:17}
\end{equation}
predicted in textures 3 and 5 of Ref.~\cite{papiro1}. In order to investigate
this relation, we
define the ratio
\begin{equation}
R_\riii \equiv \frac{1}{|V_{ub}|} \sqrt \frac{m_u}{m_t} \,.
\label{ec:18}
\end{equation}
The effects of random perturbations on this quantity can be seen in
Fig.~\ref{fig:R3}. (Recall that here $R_\riii$ is evaluated
at the scale $M_Z$. The test of this relation on the Yukawa matrices
must be done at very high
energies, using the RG equations for the SM or the SM extension under
consideration to evolve the Yukawa couplings.)

\begin{figure}[htb]
\begin{center}
\begin{tabular}{cc}
\mbox{\epsfig{file=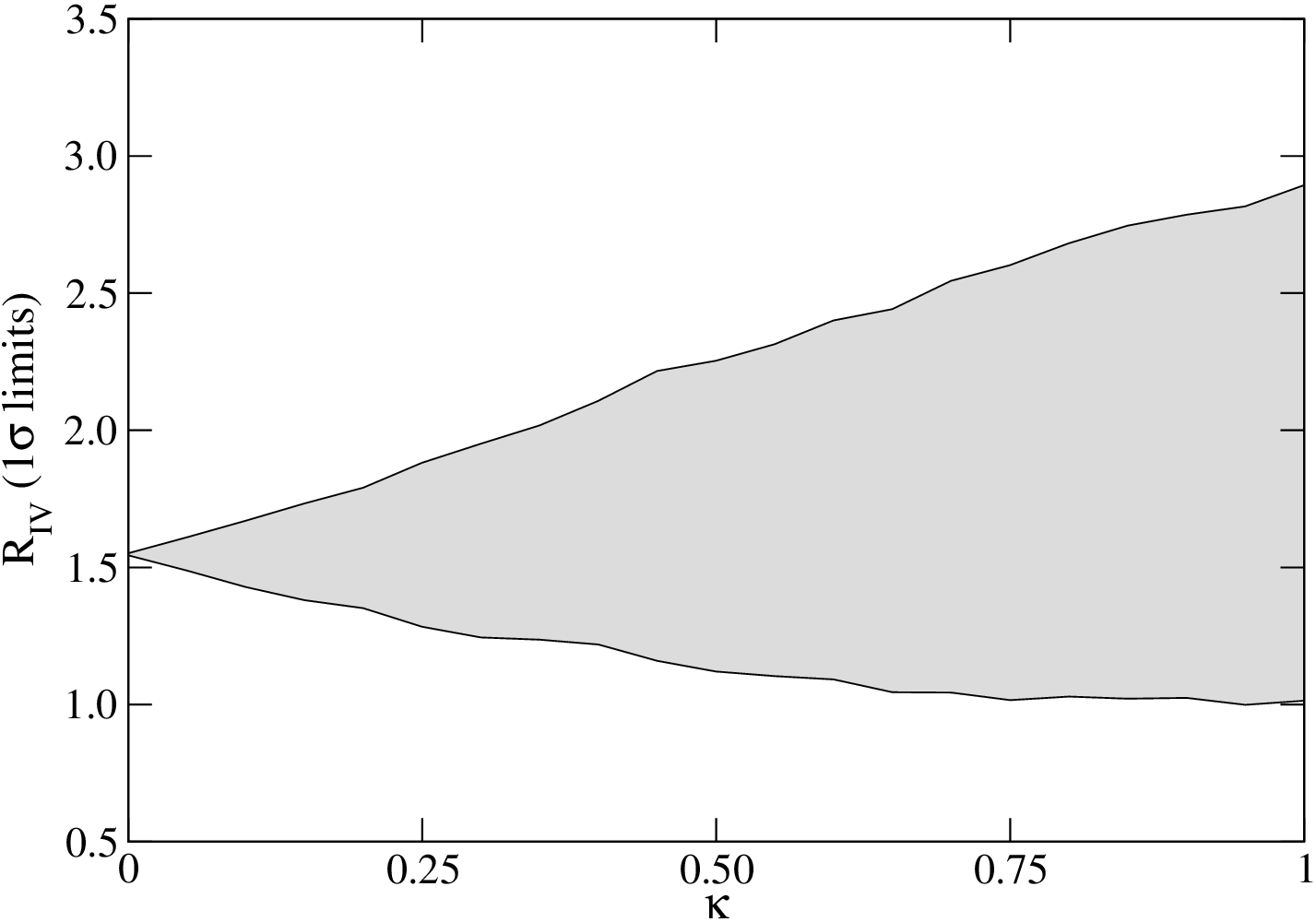,width=7cm,clip=}} &
\raisebox{-0.16cm}{\epsfig{file=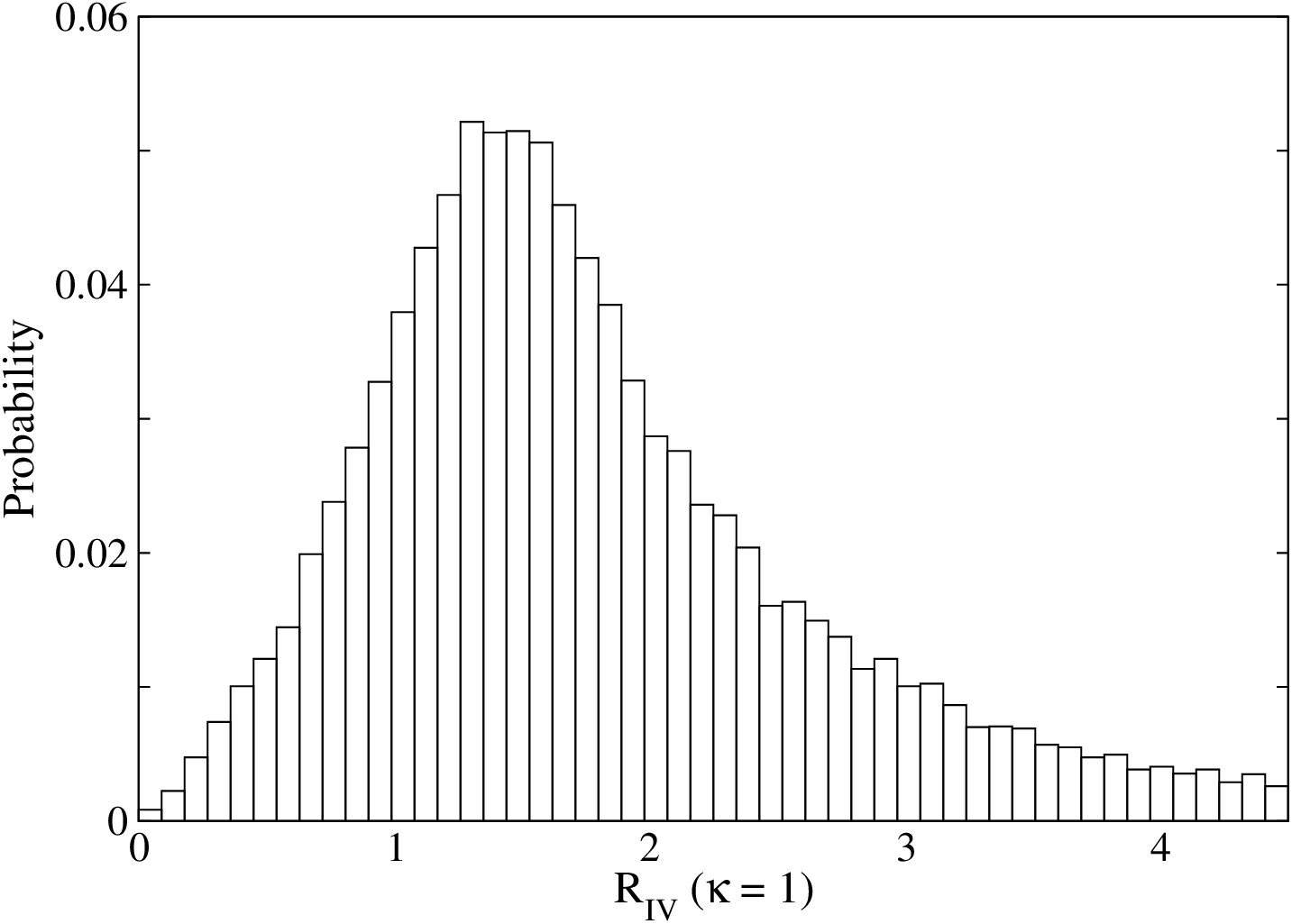,width=7.02cm,clip=}}
\end{tabular}
\end{center}
\caption{Effect of the random perturbations on the ratio $R_\riv$, defined in
the text.
\label{fig:R4} }
\end{figure}

The plot in Fig.~\ref{fig:R3} (left) shows that the perturbations change this
ratio from an initial value $(R_\riii)_0 = 0.83$ to an interval between $0.55$
and $1.56$. This implies that the corrections to the Yukawa matrices might
hide an underlying relation given by Eq.~(\ref{ec:17}). Relation IV is
\cite{papiro10}
\begin{equation}
|V_{cb}| = \sqrt \frac{m_c}{m_t} \,.
\label{ec:19}
\end{equation}
Correspondingly, we define
\begin{equation}
R_\riv \equiv \frac{1}{|V_{cb}|} \sqrt \frac{m_c}{m_t} \,.
\label{ec:19b}
\end{equation}
This quantity is plotted in Fig.~\ref{fig:R4}. Since the ratios $m_u/m_c$ and
$|V_{ub}/V_{cb}|$ are not modified by the random perturbations, the behaviour
is similar to $R_\riii$, up to a global factor.

\begin{figure}[htb]
\begin{center}
\begin{tabular}{cc}
\mbox{\epsfig{file=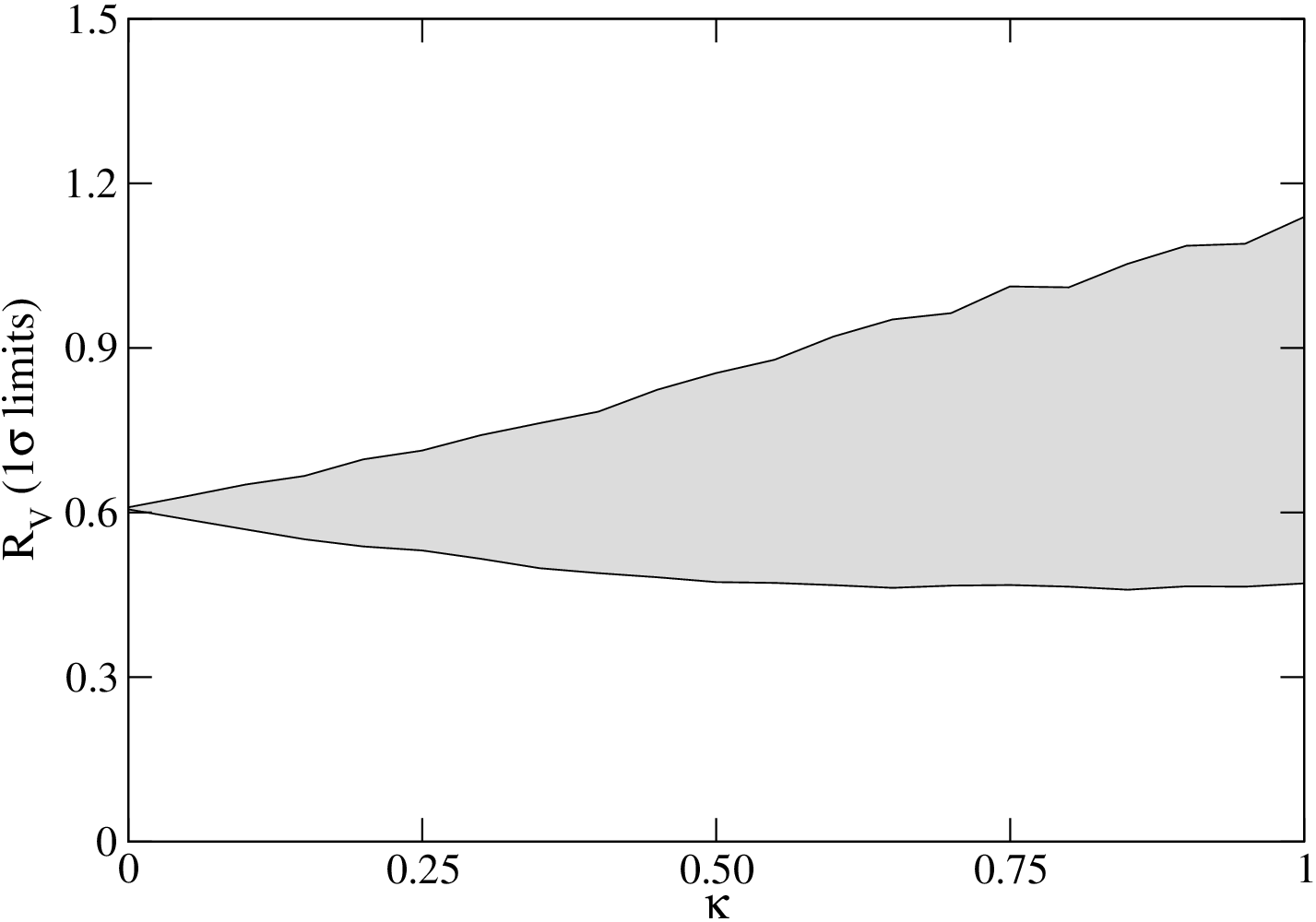,width=7cm,clip=}} &
\raisebox{-0.15cm}{\epsfig{file=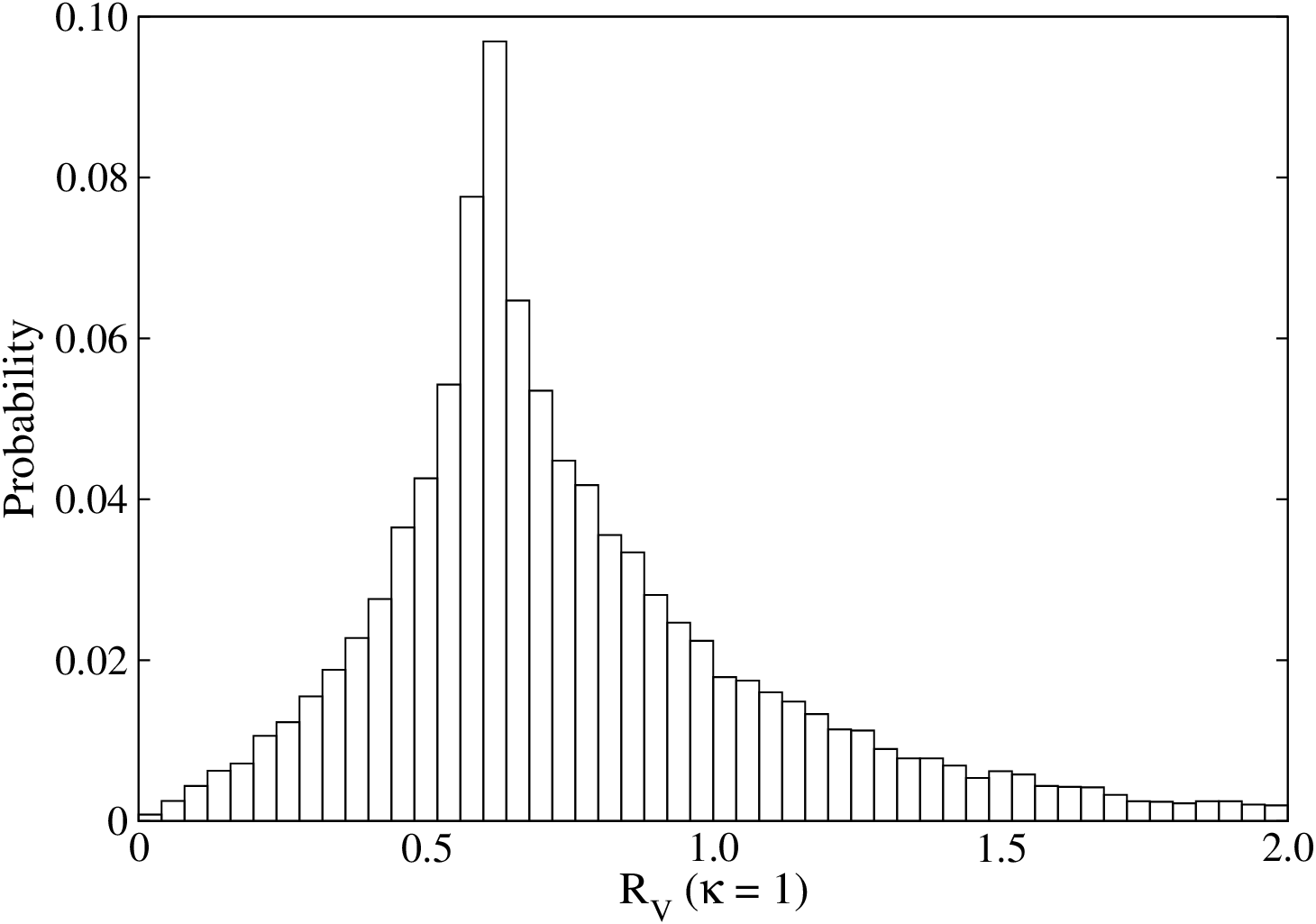,width=7.11cm,clip=}}
\end{tabular}
\end{center}
\caption{Effect of the random perturbations on the ratio $R_\rv$, defined in
the text.
\label{fig:R5}}
\end{figure}

\begin{figure}[htb]
\begin{center}
\begin{tabular}{cc}
\mbox{\epsfig{file=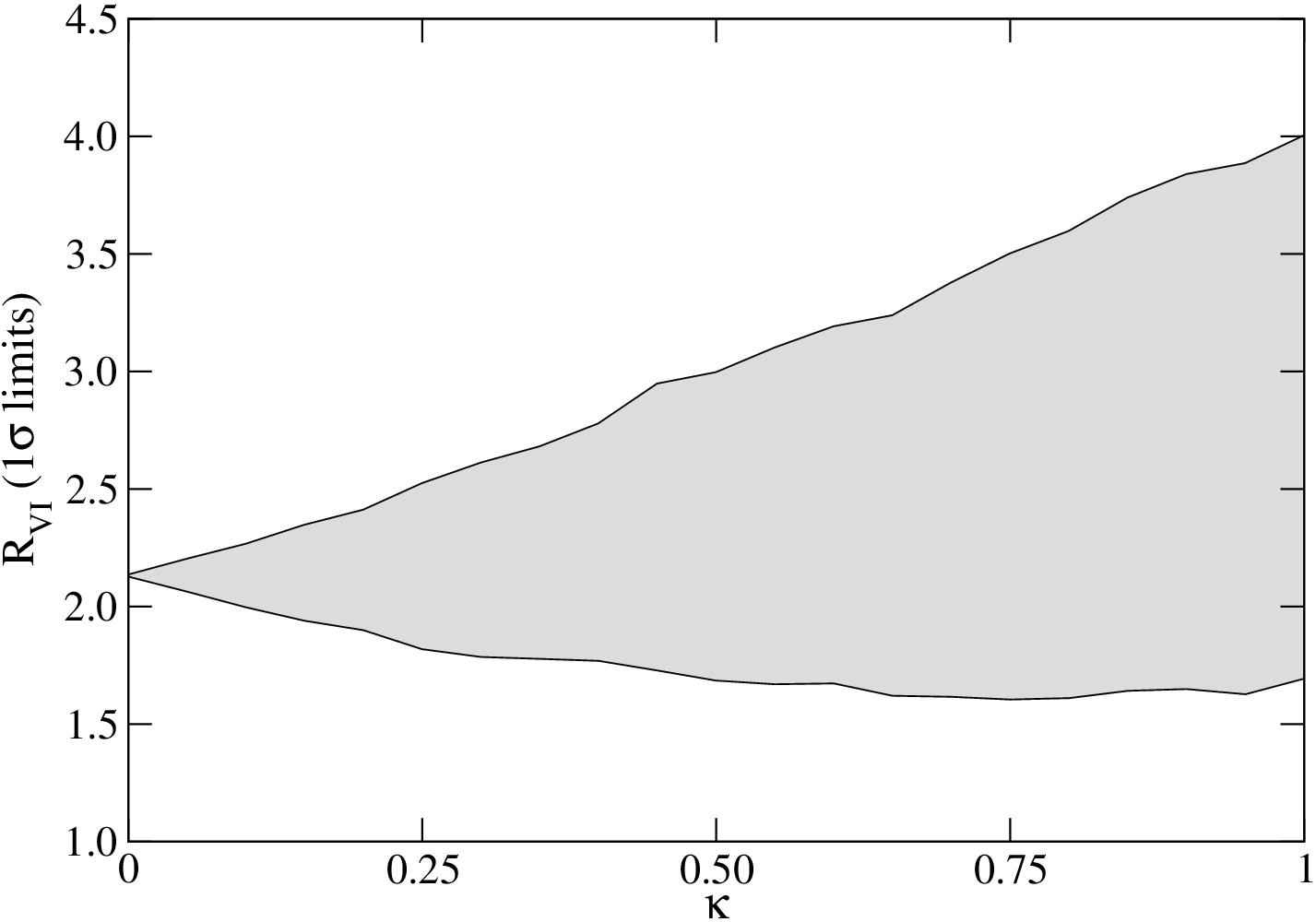,width=7cm,clip=}} &
\raisebox{-0.13cm}{\epsfig{file=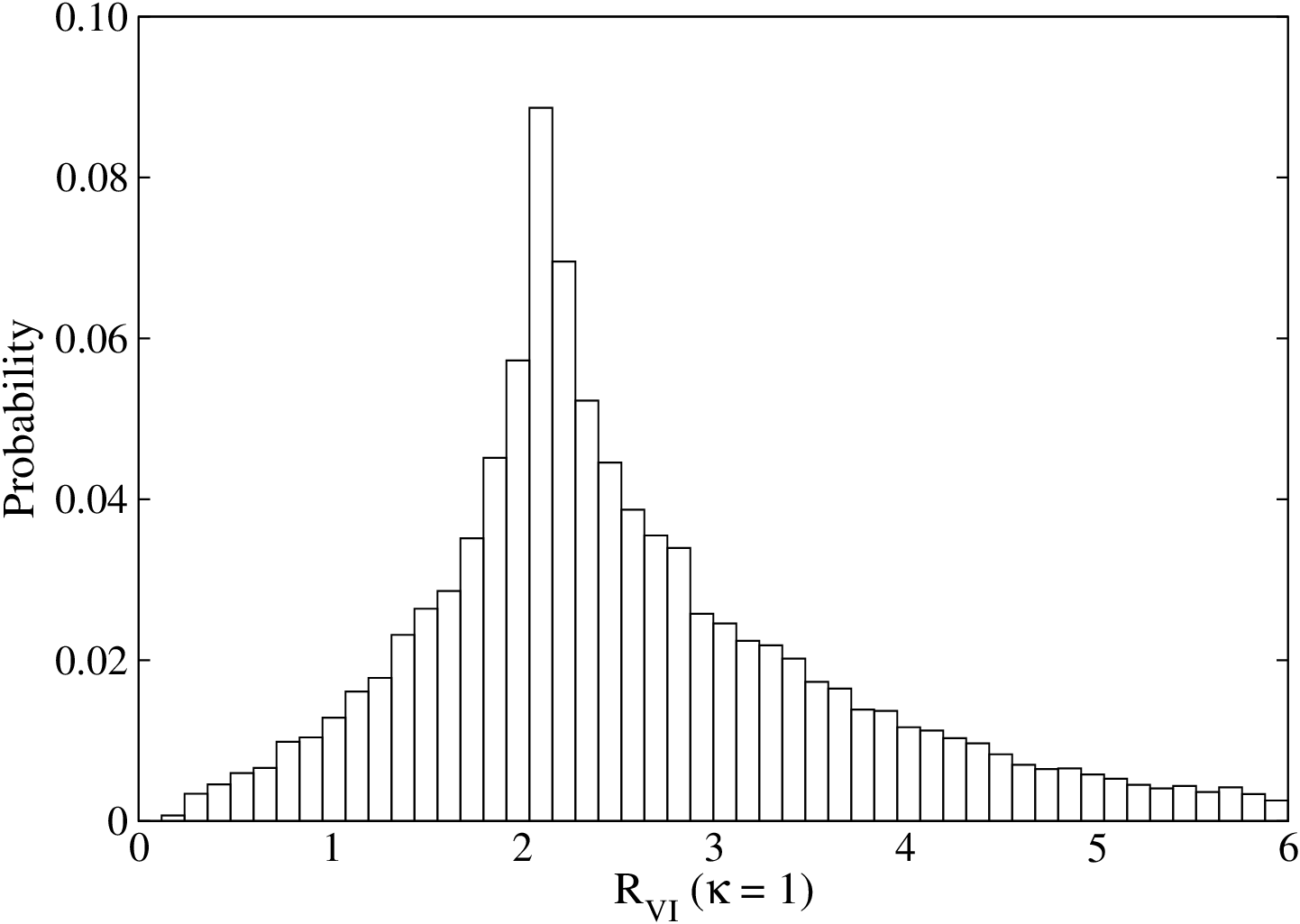,width=7.05cm,clip=}}
\end{tabular}
\end{center}
\caption{Effect of the random perturbations on the ratio $R_\rvi$, defined in
the text.
\label{fig:R6}}
\end{figure}

Relations V and VI \cite{papiro12} are very similar and involve only
quark masses:
\begin{equation}
\frac{m_u}{m_c} = \frac{m_c}{m_t} ~,~~~
\frac{m_d}{m_s} = \frac{m_s}{m_b} \,.
\label{ec:20}
\end{equation}
The impact of the ``anarchy'' on them can be investigated with the
analysis of the ratios
\begin{equation}
R_\rv \equiv \frac{m_u m_t}{m_c^2} ~,~~~
R_\rvi \equiv \frac{m_d m_b}{m_s^2} \,,
\label{ec:21}
\end{equation}
which are plotted in Figs.~\ref{fig:R5} and \ref{fig:R6}, respectively. These
quantities are (up to global factors) the inverse of the ratios $m_c/m_t$ and
$m_s/m_b$ studied above.

Relation VII \cite{papiro13,papiro14} is analogous to Relation $\rii$ in
Eq.~(\ref{ec:16}), 
\begin{equation}
\left| \frac{V_{td}}{V_{ts}} \right| = \sqrt \frac{m_d}{m_s} \,.
\label{ec:22}
\end{equation}
Since the ratio $|V_{ub}/V_{cb}|$ and the phase $\delta$ do not change with
the perturbations, Relation VII remains inaltered as well. Relation
VIII is \cite{papiro13}
\begin{equation}
|V_{cb}| = \sqrt 2 \, \frac{m_s}{m_b} \,.
\label{ec:23}
\end{equation}
In principle, both sides of this equality change when the random perturbations
are added to the Yukawa matrices. However, the changes are correlated and the 
relation is not affected. The same happens with Relation IX \cite{papiro13},
\begin{equation}
|V_{ub}| = \sqrt \frac{m_d m_s}{2 m_b^2} \,.
\label{ec:24}
\end{equation}
The non-trivial invariance of these relations under the ``anarchy'' is due to
the formal structure adopted for the latter, Eqs.~(\ref{ec:6}). 

As remarked at the beginning of this Section, the test of Relations I--IX must
be done at very high energies, where the flavour symmetry originating the
Yukawa patterns is not broken. In this context,
we would like to point out that the test of
Relations I, II, VII, VIII and IX, as well as other relations invariant under
random perturbations, can provide a better
way to discriminate between different models of Yukawa patterns. On the other
hand, Relations III, IV, V and VI can be influenced by radiative corrections
from new physics and hence their observation might be unclear.

\section{Enhancement of the hierarchy}
\label{sec:5}

We want to investigate within our framework to what extent the
``anarchy'' might contribute to (or even generate) the observed hierarchy of
quark masses and mixings. The guideline of our analysis is the following:
we reduce the hierarchy in the
parameters of the Lagrangian and then calculate the probability that the
hierarchy present in the SM is reproduced with the random perturbations.
We have studied three examples, corresponding to an enhancement of the
hierarchy in the up-type quark masses, down-type quark masses and CKM mixing
angles, respectively.
\begin{enumerate}
\item In the first case, we use the same parameters as in Section \ref{sec:3}
but replace the top mass by $m_t'=m_t/2$. In this situation, the ratio of masses
$m_c/m_t$ is larger by a factor of two, $(m_c/m'_t)_0 = 7.7 \times 10^{-3}$
(that is, the hierarchy is reduced).
Setting $\kappa=1$, we generate a sample of 10000 random Yukawa matrices.
A good estimate of the likelihood to enhance the hierarchy is given
by the probability that the ratio $m_c/m'_t$ after perturbations is smaller or
equal than the value $m_c/m_t$ obtained with the true top mass. This probability
is only $8.3 \times 10^{-3}$.
\item In the second example, we use the standard parameters but replace the
bottom mass by $m_b' = m_b/2$. With the same procedure, we find that the
probability that the ratio $m_s/m_b'$ after random perturbations is smaller
or equal to the SM value is
$0.12$, more than one order of magnitude larger than in the previous case.
\item In the third example, we use the SM values of the masses, the mixing
angle $|V_{us}|$ and the phase $\delta$,
multiplying $|V_{ub}|$ and $|V_{cb}|$ by two. The probability
that the ratio $|V_{cb}/V_{us}|$ after random perturbations is equal or
smaller than the SM value is $0.12$.
\end{enumerate}
With these examples we see that the observed large hierarchy between the
second and
third generations is not an effect of the ``anarchy'' (as discussed in
Ref.~\cite{papiro16}), though for down-type quarks it can receive some
enhancements from unknown corrections.
The two mixing angles $V_{ub}$, $V_{cb}$ can also be reduced by the
corrections.

\section{Summary}
\label{sec:6}

We have introduced a basis-independent parameterisation in order to describe the
effect of unknown corrections from new physics (``anarchy'') in
the quark and lepton mass
matrices. With this parameterisation we have explored the stability of some
properties of the quark Yukawa matrices against unknown corrections. This has
been done with the addition of random perturbations to these matrices and a
statistical analysis of the effect of the perturbations.

We have shown that the quark mass hierarchies, namely the ratios $m_u/m_c$,
$m_c/m_t$, $m_d/m_s$, $m_s/m_b$ and $m_b/m_t$ are hardly affected by the random
perturbations. Of these quantities, $m_u/m_c$ and $m_d/m_s$ remain constant for
most of the values of the random parameters. The rest exhibit deviations that
in average lead to an enhancement of the original hierarchy, but without
modifying it significantly. We have also
analysed the effect in the mixing angles, concluding that neither $|V_{us}|$,
nor the ratio $|V_{ub}/V_{cb}|$, nor the phase $\delta$ change appreciably
with the random
perturbations. The mixing angles $|V_{ub}|$ and $|V_{cb}|$ show deviations
but still preserving the CKM hierarchy $|V_{us}| \gg |V_{cb}| \gg |V_{ub}|$.
For some fine-tuned values of the random parameters, the strong hierarchy of
masses and mixing angles is
removed, but this only happens in a extremely small subset of the sample.
We have also found that when the size of the random perturbations (the $\kappa$
parameter) is increased, the average effects do not grow linearly but their
increase rate slows down.

We have selected nine simple flavour relations among quark masses and CKM
mixing angles, predicted by several models in the literature,
discussing the effect of the ``anarchy'' on them. We have identified
four relations which are affected by the random perturbations. If these
relations
are fulfilled by the Yukawa matrices, as some theoretical models predict
\cite{papiro1,papiro10,papiro12},
the phenomenological observation may be jeopardised by the ``anarchy'':
some flavour properties in the Lagrangian might not be apparent
and the ``anarchy'' might blur or hide an underlying flavour relation.
On the other hand, the remaining five relations discussed are not altered by
the
``anarchy'' and hence their analysis could provide a cleaner insight into the
structure of the Yukawa matrices.

Finally, we have used our framework to demonstrate that the possibility
that unknown corrections give large contributions to the hierarchy of masses
and mixings is very unlikely, at it has been pointed out before in the
literature. We have shown that if, for instance, one enhances
the ratio $m_c/m_t$ in the Yukawa matrices in the Lagrangian (thus reducing the
hierarchy), the probability
that the corrections bring it down to the SM value is very small.

\vspace{1cm}
\noindent
{\Large \bf Acknowledgements}

\vspace{0.4cm} \noindent
I thank F. del \'Aguila and A. M. Teixeira for a careful reading of the
manuscript.
This work has been supported by the European Community's Human Potential
Programme under contract HTRN--CT--2000--00149 Physics at Colliders.


\begin{thebibliography}{99}
\bibitem{papiro1}
P. Ramond, R. G. Roberts and G. G. Ross, Nucl. Phys. B {\bf 406} (1993) 19

\bibitem{papiro2}
H. Harari, H. Haut and J. Weyers, Phys.\ Lett.\ B {\bf 78} (1978) 459;
for a recent analysis see
G. C. Branco, M. E. G\'omez, S. Khalil and A. M. Teixeira, hep-ph/0204136

\bibitem{papiro15}
H.~Fritzsch, Phys. Lett. {\bf 73B} (1978) 317

\bibitem{papiro15b}
B. Stech, Phys. Lett. B {\bf 130} (1983) 189

\bibitem{papiro16}
C. D. Froggatt and H. B. Nielsen, Nucl. Phys.B {\bf 147} (1979) 277

\bibitem{papiro11}
R. Rosenfeld and J. L. Rosner, Phys. Lett. B {\bf 516} (2001) 408

\bibitem{papiro17}
S. Fukuda {\em et al.}  [Super-Kamiokande Collaboration], Phys. Rev. Lett.
{\bf 86} (2001) 5656

\bibitem{papiro3}
L. J. Hall, H. Murayama and N. Weiner, Phys. Rev. Lett. {\bf 84} (2000) 2572;
N. Haba and H. Murayama, Phys. Rev.D {\bf 63} (2001) 053010

\bibitem{papiro3b}
G. Altarelli, F. Feruglio and I. Masina, hep-ph/0210342

\bibitem{papiro4}
T. P. Cheng, E. Eichten and L. F. Li, Phys. Rev. D {\bf 9} (1974) 2259
\bibitem{papiro4b}
H. Arason, D. J. Casta\~no, B. Keszthelyi, S. Mikaelian, E. J. Piard, P. Ramond
and B. D. Wright, Phys. Rev. D {\bf 46} (1992) 3945

\bibitem{papiro5}
F. del Aguila, J. A. Aguilar-Saavedra and M. Zralek, Comput. Phys. Commun.
{\bf 100} (1997) 231

\bibitem{papiro6}
J. A. Aguilar-Saavedra {\em et al.}, in preparation

\bibitem{papiro7}
K. Hagiwara {\em et al.}, Particle Data Group, Phys. Rev. D {\bf 66}
(2002) 010001 

\bibitem{papiro8}
J. A. Aguilar-Saavedra and M. Masip, Phys. Rev. D {\bf 54} (1996) 6903

\bibitem{papiro9}
R. Gatto, G. Sartori and M. Tonin, Phys. Lett. B {\bf 28} (1968) 128
\bibitem{papiro9b}
R. J. Oakes, Phys. Lett. B {\bf 29} (1969) 683

\bibitem{papiro10}
S. Dimopoulos, L. J. Hall and S. Raby, Phys. Rev. Lett. {\bf 68} (1992) 1984;
Phys. Rev. D {\bf 45} (1992) 4192

\bibitem{papiro12}
G. C. Branco, D. Emmanuel-Costa and R. Gonz\'alez Felipe, Phys. Lett. B
{\bf 483} (2000) 87

\bibitem{papiro13}
G. C. Branco, D. Emmanuel-Costa and J. I. Silva-Marcos, Phys. Rev. D {\bf 56}
(1997) 107

\bibitem{papiro14}
R. Barbieri, L. J. Hall and A. Romanino, Nucl. Phys. B {\bf 551} (1999) 93

\end{thebibliography}
\end{document}